\documentclass[12pt, draftcls, one column]{IEEEtran}
\usepackage{cite}
\usepackage[cmex10]{amsmath}
\usepackage{amsfonts}
\usepackage{mathrsfs}
\usepackage{xcolor}
\usepackage{bm}
\usepackage{hyperref}
\hypersetup{dvipdfm}

\usepackage[dvips, final]{graphicx}
\graphicspath{{./}}
\DeclareGraphicsExtensions{.eps}
\newcommand{\figwidth}{.7\linewidth}

\usepackage{algorithmic}
\usepackage{array}

\usepackage{mdwmath}
\usepackage{mdwtab}

\usepackage{eqparbox}
\usepackage[tight,footnotesize]{subfigure}

\begin{document}
\title{Monobit Digital Receivers for QPSK: Design, Performance and Impact of IQ Imbalances}

\author{Zhiyong~Wang,~\IEEEmembership{Student Member,~IEEE,}
        Huarui~Yin,~\IEEEmembership{Member,~IEEE,}
        Wenyi~Zhang,~\IEEEmembership{Senior Member,~IEEE,}
        and~Guo~Wei
\thanks{The authors are with Department of Electronic Engineering and Information Science, University of Science and Technology of China, Hefei, China. (e-mail: wzy2@mail.ustc.edu.cn, \{yhr, wenyizha, wei\}@ustc.edu.cn). Any comments should be addressed to Huarui Yin.}

\thanks{This research has been funded in part by the National Science Foundation of China under Grant No. 60802008, the National High Technology Research Development Program of China (863 Program) under Grant No. 2011AA010201, the National Basic Research Program of China (973 Program) through grant 2012CB316004, and the 100 Talents Program of Chinese Academy of Sciences.}
}



\maketitle

\begin{abstract}
Future communication system requires large bandwidths to achieve high data rates, thus rendering analog-to-digital conversion (ADC) a bottleneck due to its high power consumption. In this paper, we consider monobit receivers for QPSK. The optimal monobit receiver under Nyquist sampling is obtained and its performance is analyzed. Then, a suboptimal but low-complexity receiver is proposed. The effect of imbalances between In-phase (I) and Quadrature (Q) branches is carefully examined. To combat the performance loss due to IQ imbalances, monobit receivers based on double training sequences and eight-sector phase quantization \cite{DBLP:journals/corr/abs-1112-4811} are proposed. Numerical simulations show that the low-complexity suboptimal receiver suffers 3dB signal-to-noise-ratio (SNR) loss in additive white Gaussian noise (AWGN) channels and only 1dB SNR loss in multipath channels compared with matched-filter monobit receiver with perfect channel state information (CSI). It is further demonstrated that the amplitude imbalance has essentially no effect on monobit receivers. In AWGN channels, receivers based on double training sequences can efficiently compensate for the SNR loss without complexity increase, while receivers with eight-sector phase quantization can almost completely eliminate the SNR loss caused by IQ imbalances. In dense multipath channels, the effect of imbalances on monobit receivers is slight.
\end{abstract}

\newpage


\IEEEpeerreviewmaketitle


\section{Introduction}
\IEEEPARstart{I}{n} the future, communication systems are expected to provide high data rates up to several Gbits/sec. To achieve this goal, extremely large bandwidths are needed. For instance, ultra-wideband (UWB) communication occupies more than 1 GHz spectrum. Communication in the 60 GHz band \cite{978061} takes up even more. Due to the significant large bandwidths, it is a huge challenge to design a sophisticated digital receiver with affordable implementation complexity and cost.   

When the received signals of high-rate large-bandwidth systems are processed digitally, analog-to-digital converter (ADC) becomes a key bottleneck. Since the power consumption of ADC is proportional to $2^b$, where $b$ is the bit width of ADC \cite{1568684}, high-speed high-resolution ADC is power-hungry and costly. Therefore, monobit ADC has attracted significant attention of late, both in view of receiver design, e.g., \cite{1188907}, \cite{1512088}, and in view of information-theoretic aspects, e.g. \cite{WZhang}, \cite{JSingh}. As it may be easily realized by a comparator, the monobit ADC can reach tens of Gsps sampling rate with very low power consumption; see, e.g., \cite{4672032}.

There are two approaches for high-rate large-bandwidth systems: single-carrier and orthogonal frequency division multiplexing (OFDM). Single-carrier systems use pulses with short duration to occupy a large bandwidth, leading to signal sparsity in time domain. Such sparsity may not be present in OFDM systems where signals in time domain are often seen as Gaussian processes. We focus on monobit reception for single-carrier systems in this paper.

The monobit ADC is particularly suitable for UWB communication using impulse radio (IR), whose traditional reception methods mainly include coherent receiver, e.g., \cite{1097830}, autocorrelation based receiver, e.g. \cite{1267846}, \cite{1542577},  and noncoherent receiver, e.g. \cite{1388736}. After monobit sampling was introduced, a matched-filter based receiver was proposed in \cite{1512088} for BPSK modulation. However, it is not optimal under Nyquist sampling as shown in \cite{5474634}, and moreover, the requirement for full resolution (FR) perfect received waveform makes it difficult to implement.

The optimal monobit receiver for BPSK was proposed in \cite{5474634}, which turns out to take the form of a linear combiner. By a Taylor expansion of the optimal weights, a suboptimal receiver was also obtained in \cite{5474634}, which is easy to implement and only incurs a slight performance loss compared with the idealized one in \cite{1512088}, even without the channel state information (CSI).

In order to achieve higher data rates, higher-order modulations such as quadrature-phase-shift-keying (QPSK) or quadrature-amplitude-modulation (QAM) are considered. For such modulations, the achievable rate with uniform low-resolution quantization is given by \cite{5503146}, while the capacity of noncoherent additive white Gaussian noise (AWGN) channel with ``phase-quantization'' is calculated in \cite{DBLP:journals/corr/abs-1112-4811}. However, the problem of receiver design is not discussed in those works.

In this paper, we study the design and performance of digital receivers for QPSK with monobit Nyquist rate sampling. First, the maximum likelihood (ML) receiver is derived, and its performance is analyzed in the form of deflection ratio \cite{395235}. Then, the main ideas in \cite{5474634} are extended herein to obtain a suboptimal receiver for QPSK. The effect of phase offset between the transmitter (Tx) and the receiver (Rx) is investigated. Compared with the matched-filter based monobit receiver with perfect CSI, the suboptimal receiver without any prior CSI has 3dB signal-to-noise-ratio (SNR) loss in AWGN channel and only 1dB SNR loss in multipath channels, when bit error rate (BER) is around $10^{-3}$. It is also observed that the suboptimal monobit receiver has only 3dB SNR loss even compared with the full-resolution matched-filter (FRMF) in dense multipath channels.

A limiting factor of practical systems is the imbalances, including amplitude and phase imbalances, between the In-phase (I) and Quadrature (Q) branches when received radio-frequency (RF) signal is down-converted to baseband \cite{Razavi:1998:RM:274817}. For its advantages in cost, area and power, the direct conversion receiver considered in this paper tends to be adopted in future RF systems. However, it is affected more seriously by the IQ imbalances than heterodyne receiver \cite{482187}. Due to monobit sampling, signals distorted by IQ imbalances are much more difficult to detect, compared with those with high-resolution ADC which has been well investigated. Besides, we are not aware of any published work dealing with the IQ imbalances under low-resolution sampling.

We investigate the effect of IQ imbalances on monobit receivers. Monobit receivers based on double training sequences are proposed to mitigate the SNR loss caused by IQ imbalances, without increasing the implementation complexity. To further improve monobit reception under IQ imbalances, a eight-sector phase quantization scheme proposed in \cite{DBLP:journals/corr/abs-1112-4811} is employed and the corresponding design of monobit receiver is obtained, at the cost of doubling the complexity in the digital domain. It is shown that the amplitude imbalance has essentially no impact on monobit receivers. It is demonstrated that the monobit receiver with eight-sector phase quantization can almost completely eliminate the SNR loss caused by IQ imbalances in AWGN and sparse multipath channels. Thanks to the diversity offered by dense multipath, monobit receivers based on traditional architecture are more desirable in dense multipath channels compared with receivers with eight-sector phase quantization, for their slight performance loss but half the implementation complexity.

The remaining part of the paper is organized as follows. The system model is presented in Section \ref{SystemModel}. The optimal and suboptimal monobit receivers without IQ imbalance are given in Section \ref{RecBalance}, along with discussion on several important implementation issues such as channel estimation, effect of phase offset, and interface with error-control decoder. Section \ref{RecImbalance} discusses the effect of IQ imbalances and proposes monobit receivers based double training sequences or eight-sector phase quantization to combat the performance degradation. Application of the proposed receivers to UWB and 60GHz systems with numerical results is provided in Section \ref{NumericalResults}. Finally, Section \ref{Conclusions} concludes the paper.

\section{System Model And Receiver Architecture}\label{SystemModel}
\begin{figure}
\centering
\includegraphics[width=\textwidth]{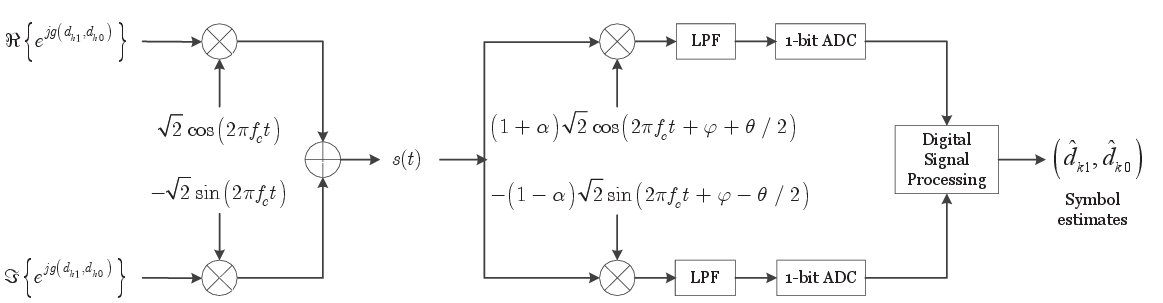}
\caption{Monobit digital receiver for QPSK with phase offset $\varphi$, amplitude imbalance $\alpha$ and phase imbalance $\theta$}
\label{Archit}
\end{figure}

The monobit digital receiver we study is depicted in the block diagram in Figure \ref{Archit}. The received baseband signal is composed of I and Q components, both of which are first filtered by an ideal low pass filter (LPF) of bandwidth $B$, then sampled and quantized by a monobit ADC at Nyquist rate $2B$. The digitized signals are processed by the digital signal processing (DSP) unit.

In this paper, the Gray-coded QPSK modulation is employed. Thus, the transmitted signal can be written as
\begin{equation}
s\left(t\right) = \sqrt{2} \mathfrak{R}\left\{ \sum\limits_{k=0}^{\infty}e^{j\left(g\left(d_{k1},d_{k0}\right)+2\pi f_ct\right)}p_\text{tr}\left(t-kT_s\right)\right\},
\end{equation}
where $k$ is the symbol index, $T_s$ is the symbol duration, $f_c$ is the carrier frequency, $d_{k0},d_{k1} \in \{+1,-1\}$ are the binary data of the $k$th QPSK symbol, and $p_\text{tr}\left(t\right)$ is the spectral shaping pulse. The mapping of $g\left(d_{k1},d_{k0}\right)$ is according to Gray coding, that is $g\left(1,1\right)=0$, $g\left(1,-1\right)=\pi/2$, $g\left(-1,1\right)=-\pi/2$ and $g\left(-1,-1\right)=\pi$. We assume that $d_{k0}$ and $d_{k1}$ are equally likely to be $\pm{1}$, and they can be either uncoded or coded.

At the receiver, the received RF signal is first down-converted to baseband. Then, both the I and Q components of the baseband signal are filtered by an ideal LPF respectively. Define $p_\text{ref}\left(t\right)=p_\text{tr}\left(t\right)\star h\left(t\right) \star p_\text{rec}\left(t\right)$ as the reference signal, where $\star$ denotes convolution, $h\left(t\right)$ and $p_\text{rec}\left(t\right)$ are the impulse response of the channel and the LPF, respectively. The channel is modeled as a linear time-invariant system within the channel coherence time. If there is no imbalance between the I and Q branches, the filtered baseband signal is given as
\begin{equation}
r_\text{b}\left(t\right) = \sum\limits_{k=0}^{\infty}e^{j\left(g\left(d_{k1},d_{k0}\right)-\varphi\right)} p_\text{ref}\left(t-kT_s\right)+n_\text{b}\left(t\right),
\end{equation}
where $\varphi$ is the carrier phase offset between the transmitter and the receiver, and $n_\text{b}\left(t\right)$ is the baseband-equivalent Gaussian noise whose variance has been normalized to one by the LPF. The phase offset $\varphi$ is an unknown constant over a block of symbols with uniform distribution on $\left[0,2\pi\right)$. Due to this fact, rotating a constellation with arbitrary angle will result in the same reception performance. We remark that such model suits both the carrier synchronous systems\footnote{Those systems include ideal synchronous systems \cite{pll_phase_offset} and systems with frequency offset smaller than 50ppb according to synchronization standards of the NIST (National Institute of Standards and Technology, USA)\cite{nistpdf}.} and asynchronous systems\footnote{Those systems have frequency offsets about several KHz (i.e. of the order of ppm with several GHz carrier frequency), such as the receivers with digital fractional-N PLL \cite{digitalpll}.} with practical values of frequency offset\footnote{In both synchronous and asynchronous cases, the block length could be of the order of hundred to thousand, since the corresponding variation of the phase offset is small enough for monobit receivers, no matter it is caused by the frequency drift \cite{634659}\cite{adf4351} or the frequency offset.}. 

Due to analog component imperfections, the I and Q branches of the receiver usually do not have equal amplitude or exact $90^{\circ}$ phase difference, leading to the amplitude and phase imbalances. Following \cite{Razavi:1998:RM:274817}, the received baseband signal distorted by IQ imbalances can be modeled as
\begin{equation}  \label{RecSigImb}
r_\text{d}\left(t\right)=\mu r_\text{b}\left(t\right)+\upsilon r_\text{b}^*\left(t\right),
\end{equation}
where $\mu$ and $\upsilon$, characterizing the imbalance between the I and Q branches, are given by
\begin{equation} \label{imbpara}
\begin{split}
\mu = \cos\left(\theta/2\right)-\mathrm{j}\alpha \sin\left(\theta/2\right),~\text{and}~~ 
\upsilon = \alpha \cos\left(\theta/2\right)+\mathrm{j}\sin\left(\theta/2\right),
\end{split}
\end{equation}
where $\theta$ denotes the phase deviation between the I and Q branches from $90^{\circ}$, and $\alpha$ denotes the amplitude imbalance given as
\begin{equation}
\alpha = \frac{a_I-a_Q}{a_I+a_Q},
\end{equation}
where $a_I$ and $a_Q$ are the gain amplitudes on the I and Q branches, respectively. When measured in dB, the amplitude imbalance is computed as $10\log_{10}(1+\alpha)$. 

We choose the filter bandwidth $B$ and the sampling period $T$ to satisfy $T=1/\left(2B\right)=T_s/N$, so that the Nyquist-rate sampling of the filtered signal is used and every pulse is sampled by $N$ points. Within the $k$th symbol, we denote the $l$th samples of the I and Q branches as $r_{I,k,l}$ and $r_{Q,k,l}$ respectively. Then we have, with monobit ADC,
\begin{equation}
r_{I,k,l}=
\begin{cases}
1,  & r_{\text{d},I}\left(kT_s+lT\right) > 0     \\
-1, & r_{\text{d},I}\left(kT_s+lT\right) \le 0
\end{cases}
~~~~~l=0,...,N-1,
\end{equation}
and
\begin{equation}
r_{Q,k,l}=
\begin{cases}
1,  & r_{\text{d},Q}\left(kT_s+lT\right) > 0     \\
-1, & r_{\text{d},Q}\left(kT_s+lT\right) \le 0
\end{cases}
~~~~~l=0,...,N-1,
\end{equation}
where $r_{\text{d},I}\left(t\right)$ and $r_{\text{d},Q}\left(t\right)$ denote the received signals of the I and Q branches respectively. We assume that the maximum channel delay spread is significantly smaller than the symbol duration $T_s$ so that ISI is negligible.

Define $\mathbf{r}_k=\left[r_{I,k,0},r_{Q,k,0},...,r_{I,k,N-1},r_{Q,k,N-1}\right]^T$. The DSP unit of the receiver is hence to detect $d_{k0}$ and $d_{k1}$ based on $\mathbf{r}_k$.

\section{Monobit Receivers Without IQ Imbalance}\label{RecBalance}
In this section, we first derive the optimal monobit detector, assuming that there is no IQ imbalance at the receiver. However, the precise reference signal $p_\text{ref}\left(t\right)$ and phase offset $\varphi$ required by such receiver are not available in practice. Hence, we then extend the main ideas in \cite{5474634} to QPSK modulation, and obtain a suboptimal but practical monobit receiver for QPSK. The performance of these monobit receivers, the effect of phase offset, and the interface with error-control decoder are also discussed.

\subsection{Optimal Monobit Receiver}
For QPSK, $d_{k0}$ and $d_{k1}$ are equally likely to be $\pm 1$. This implies that the optimal detector, based on the digital samples $\mathbf{r}_k$, is the ML detector, which minimizes the symbol error rate (SER). Define  
\begin{equation}
\epsilon_{I,l} = Q\left(r_{\text{b},I}\left(lT\right)\right),~\text{and} ~~\epsilon_{Q,l} = Q\left(r_{\text{b},Q}\left(lT\right)\right),
\end{equation}
where $Q\left(\cdot\right)$ is the $Q$ function $Q\left(x\right)=\frac{1}{\sqrt{2\pi}} \int_x^{\infty}e^{-\frac{t^2}{2}}dt$, $r_{\text{b},I}\left(t\right)$ and $r_{\text{b},Q}\left(t\right)$ denote the I and Q branches of $r_\text{b}\left(t\right)$ respectively. The parameters $\epsilon_{I,l}$ and $\epsilon_{Q,l}$ can be viewed as the error probabilities for binary transmission of the $l$th ``chip'' $r_{\text{b},I}\left(lT\right)$ and $r_{\text{b},Q}\left(lT\right)$ through AWGN channel, respectively. The log-likelihood function of the $k$th symbol, denoted as $\Lambda_{}^{\left(\text{opt}\right)}\left(d_{k1},d_{k0}\right)$, is given by
\begin{equation} \label{OptMonR}
\begin{split}
\Lambda_{}^{\left(\text{opt}\right)}\left(d_{k1},d_{k0}\right) = \sum\limits_{l=0}^{N-1} \left\{\log\left(1+\left(\frac{d_{k1}+d_{k0}}{2}r_{I,k,l}+\frac{d_{k1}-d_{k0}}{2}r_{Q,k,l}\right) \left(1-2\epsilon_{I,l}\right)\right)\right.\\
\left.+\log\left(1-\left(\frac{d_{k1}-d_{k0}}{2}r_{I,k,l}-\frac{d_{k1}+d_{k0}}{2}r_{Q,k,l}\right)
\left(1-2\epsilon_{Q,l}\right)\right)\right\}-2N\log 2 .
\end{split}
\end{equation}
So the ML detector is
\begin{equation}  \label{MLDetector}
\left(\hat{d}_{k1},\hat{d}_{k0}\right) = \arg \max_{d_{k1},d_{k0}=\pm 1} \Lambda_{}^{\left(\text{opt}\right)}\left(d_{k1},d_{k0}\right).
\end{equation}
Note that the logarithmic operation in (\ref{OptMonR}) requires high implementation complexity even through table-lookup.  

Since the Gray coding is employed, we may assume that the BER is approximately equal to SER. Unfortunately, calculating the SER of the optimal monobit receiver is still tedious. Hence, we use deflection ratio as the performance criterion, not only for its simplicity, but also for the equivalence between the ML receiver and the optimum receiver in terms of deflection \cite{395235}.

Define $\lambda_k = \Lambda^{\left(\text{opt}\right)}\left(d_{k1}=1,d_{k0}=1\right)$ as the decision statistic of the ML detector. The deflection ratio under QPSK modulation with monobit sampling is given as \cite{395235}
\begin{equation}  \label{DefRatio}
D=\frac{\left[E\left(\lambda_k|d_{k1}=1,d_{k0}=1\right)-E\left(\lambda_k|d_{k1}=1,d_{k0}=-1\right)\right]^2} {\mathrm{var}\left(\lambda_k\right)}.
\end{equation}
After manipulations, the deflection ratio of the optimal monobit receiver is given as
\begin{equation}
D^{\left(\text{opt}\right)}=2\frac{\left[\sum\limits_{l=0}^{N-1}\left(\left(1-\epsilon_{I,l}-\epsilon_{Q,l}\right)\log\frac{1-\epsilon_{I,l} }{\epsilon_{I,l}} + \left(\epsilon_{I,l}-\epsilon_{Q,l}\right)\log\frac{1-\epsilon_{Q,l}} {\epsilon_{Q,l}}\right)\right]^2}{\sum\limits_{l=0}^{N-1}\left(\epsilon_{I,l}\left(1-\epsilon_{I,l}\right)+ \epsilon_{Q,l}\left(1-\epsilon_{Q,l}\right)\right)\left(\log^2\frac{1-\epsilon_{I,l}}{\epsilon_{I,l}} + \log^2\frac{1-\epsilon_{Q,l}}{\epsilon_{Q,l}}\right)}.
\end{equation}

From another perspective, the decision statistic $\lambda_k$ can be treated as a Gaussian random variable using a central limit argument, when $N$ is large. Thus, the BER performance can be approximately estimated as $Q\left(\sqrt{D}\right)$, which makes the deflection ratio a performance indicator. It can thus be observed that the reception performance will be good if the deflection ratio is large.

With the information of received reference waveform $p_\text{ref}\left(t\right)$ and phase offset $\varphi$, we can evaluate the deflection ratio of the optimal monobit receiver to obtain an assessment about its performance.

\subsection{Suboptimal Monobit Receiver}
To get the knowledge of $p_\text{ref}\left(t\right)$, a reference estimator based on training sequence was proposed in \cite{4712729} for BPSK transmission. However, that estimator requires a large lookup table and exhaustive search. The method proposed in \cite{5474634} only employs bit-level addition and shift operations to recover the reference signal from monobit sampling results, and thus can be conveniently realized online. Herein, the main ideas to obtain the suboptimal monobit receiver in \cite{5474634} are extended to QPSK modulation.
 
The first technique used is the Taylor's expansion. Define $w_{I,l}=1-2\epsilon_{I,l}$ and $w_{Q,l}=1-2\epsilon_{Q,l}$. When SNR is small, $\epsilon_{I,l}\approx 0.5$ and $\epsilon_{Q,l}\approx 0.5$, which lead to $w_{I,l}\approx 0$ and $w_{Q,l}\approx 0$. Thus, we can perform a first-order Taylor's expansion of the $\log$ functions in (\ref{OptMonR}), according to $\log\left(1+x\right)\approx 1+x$ when $|x|\ll 1$. Ignoring the constant $-2N\log 2$, we obtain the linear approximation of (\ref{OptMonR}) as
\begin{equation} \label{SubMonRv}
\begin{split}
\Lambda_{}\left(d_{k1},d_{k0}\right) = & \sum\limits_{l=0}^{N-1} \left\{  w_{I,l}  \left(\frac{d_{k1}+d_{k0}}{2}r_{I,k,l}+ \frac{d_{k1}-d_{k0}}{2}r_{Q,k,l}\right) \right. \\ &\left. -w_{Q,l} \left(\frac{d_{k1}-d_{k0}}{2}r_{I,k,l}-\frac{d_{k1}+d_{k0}}{2}r_{Q,k,l} \right) \right\}. 
\end{split}
\end{equation}
Compared with the ML detector, the suboptimal detector of (\ref{SubMonRv}) has greatly reduced implementation complexity. 

Due to the monobit quantization, the receiver cannot obtain the perfect reference signal $p_\text{ref}\left(t\right)$ and phase offset $\varphi$, or equivalently $\epsilon_{I,l}$, $\epsilon_{Q,l}$. Hence, we need to estimate $\epsilon_{I,l}$ and $\epsilon_{Q,l}$, to further obtain $w_{I,l}$ and $w_{Q,l}$. Assume that a sequence of training symbols (say $N_t$ symbols) are used for estimation. Without loss of generality, all symbols in the training sequence are assumed to be $\left(d_{1}=1,d_{0}=1\right)$. Then the ML estimates of $w_{I,l}$ and $w_{Q,l}$ can be given as
\begin{equation}   \label{EstiWeight}
\hat{w}_{I,l}=\frac{1}{N_t}\sum\limits_{k=0}^{N_t-1}r_{I,k,l}~,~~~ \hat{w}_{Q,l}=\frac{1}{N_t}\sum\limits_{k=0}^{N_t-1}r_{Q,k,l}~,~~~0 \le l < N-1.
\end{equation}
Replacing $w_{I,l}$ and $w_{Q,l}$ in (\ref{SubMonRv}) by $\hat{w}_{I,l}$ and $\hat{w}_{Q,l}$ respectively, a suboptimal monobit receiver without CSI is realized.

It is tempting to substitute $\hat{w}_{I,l}$ and $\hat{w}_{Q,l}$ into (\ref{OptMonR}) to realize an ML receiver without CSI. However, the $\log$ operation in the ML receiver still has significantly high complexity. Furthermore, our numerical experiments show that the robustness of such receiver is poor as SNR gets large, since a small estimation error will lead to poor detection performance due to the nonlinear amplification of $\log$ function. Hence, in the following we focus on the suboptimal monobit receiver when perfect CSI is unavailable.

We remark that the iterative demodulation and removal of small-weight points in \cite{5474634} can also be applied to QPSK, to improve the reception performance of suboptimal monobit receiver.

\subsection{Performance of Suboptimal Monobit Receiver}
From (\ref{EstiWeight}), we can obtain the mean and variance of the weight $\hat{w}_{I,l}$ as
\begin{equation}
E\left\{\hat{w}_{I,l}\right\} = 1-2\epsilon_{I,l}~,~~~
\mathrm{var}\left\{\hat{w}_{I,l}\right\} =\frac{4\epsilon_{I,l}\left(1-\epsilon_{I,l}\right)}{N_t};
\end{equation}
the mean and variance of $\hat{w}_{Q,l}$ can be obtained similarly. The deflection ratio of the suboptimal monobit receiver can thus be written as
\begin{equation}   \label{DefSub}
D=\frac{\left(\sum\limits_{l=0}^{N-1}M\left(\epsilon_{I,l}\right)+M\left(\epsilon_{Q,l}\right)\right)^2}
{\sum\limits_{l=0}^{N-1}\left\{M\left(\epsilon_{I,l}\right)+M\left(\epsilon_{Q,l}\right) + 4\left( V\left(\epsilon_{I,l}\right) + V\left(\epsilon_{Q,l}\right)\right)/N_t  -0.5\left( M\left(\epsilon_{I,l}\right)+M\left(\epsilon_{Q,l}\right) \right)^2 \right\}},
\end{equation}
where $M\left(x\right)=\left(1-2x\right)^2$ and $V\left(x\right)=x\left(1-x\right)$. We remark that the deflection ratio of suboptimal monobit receiver increases with $N_t$, for given reference waveform $p_\text{ref}\left(t\right)$ and phase offset $\varphi$. 

When iterative demodulation is employed \cite{5474634}, the weights $\hat{w}_{I,l}$ and $\hat{w}_{Q,l}$ are updated in each iteration. To quantify the performance gain offered by iterative demodulation, the deflection ratio after the $n$th iteration, denoted as $D_n$, is calculated. It turns out that $D_n$ can be obtained by simply replacing $N_t$ in (\ref{DefSub}) with $N_{t,n}^{eq}$, which can be regarded as the equivalent number of training symbols, given by
\begin{equation}  
N_{t,n}^\text{eq} = 
\begin{cases}
\frac{\left(N_t+N_d-N_{e,n-1}\right)^2}{N_t+N_d}, &n\ge 1 \\
N_t, &n=0
\end{cases},
\end{equation}
where $N_d$ is the number of data symbols and $N_{e,n-1}$ denotes the number of decision errors after $\left(n-1\right)$th iteration.

\subsection{Effect of Phase Offset}
From (\ref{DefSub}), it can be observed that the deflection ratio only depends on $\epsilon_{I,l}$ and $\epsilon_{Q,l}$ when $N_{t,n}^\text{eq}$ is fixed. For a given reference signal $p_\text{ref}\left(t\right)$, the parameters $\epsilon_{I,l}$ and $\epsilon_{Q,l}$ are solely determined by the phase offset $\varphi$, which is constant but unknown to the receiver. As a result, the reception performance is affected by the phase offset.

\begin{figure}
\centering
\includegraphics[width=\figwidth]{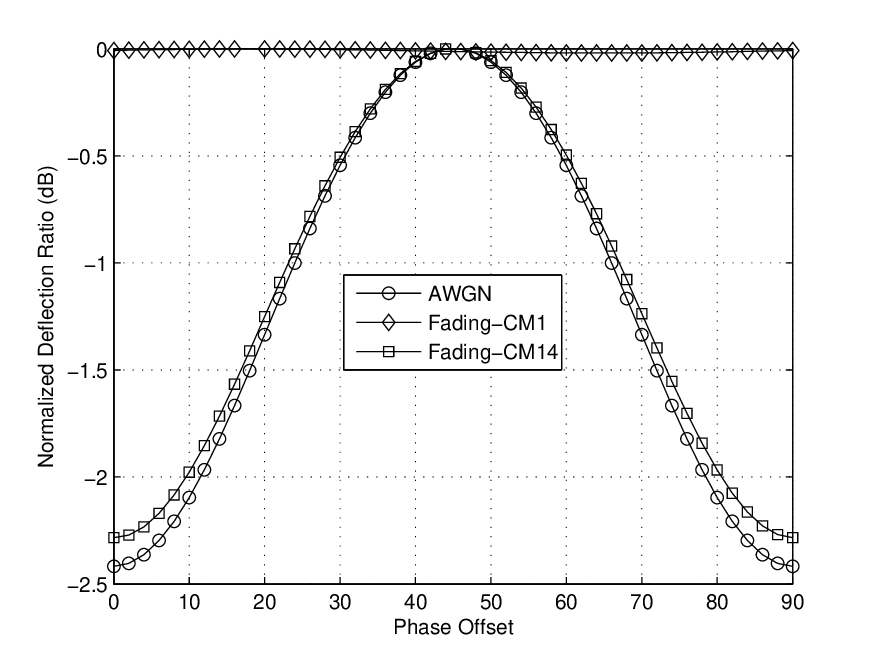}
\caption{Normalized deflection ratios for different values of phase offset $\varphi$}
\label{DefRatioEps}
\end{figure}

The deflection ratios in AWGN and multipath channels, for different values of $\varphi\in \left[0,90^{\circ}\right]$, are presented in Figure \ref{DefRatioEps}. For dense multipath channels, we use CM1 channel model \cite{chanReport} of UWB, which describes a line-of-sight (LOS) scenario with a distance between transmitter and receiver of less than 4m. For sparse multipath case, the CM14 channel model \cite{TG3c} in 60 GHz band is used. It is a two-path Saleh-Valenzuela (TSV) model for LOS scenario in residential environment, with antenna beam-widths of Tx-$15^{\circ}$ and Rx-$15^{\circ}$. It is observed that the deflection ratio will be larger if the amplitudes of the I and Q branches are closer to each other, such as the situation in AWGN and sparse multipath channels when $\varphi=45^{\circ}$. The deflection ratio decreases significantly when the amplitude difference between the two branches is large, e.g. the cases in AWGN and sparse multipath channels when $\varphi=0^{\circ}$ or $\varphi=90^{\circ}$. Fortunately, the impact of phase offset is much weaker in dense multipath channels, owing to the diversity offered by multipath and the consequent phase diversity. From this point of view, the suboptimal monobit receiver is more suitable for dense multipath channels. 

\subsection{Interface with Error-Control Decoder}
In the coded case, messages in the form of log-likelihood ratio (LLR) need to be fed to the decoder, as well as exchanged inside iterative decoders such as turbo coding. Note that $\Lambda^{\left(\text{opt}\right)}\left(d_{k1},d_{k0}\right) \approx \Lambda\left(d_{k1},d_{k0}\right)-2N\log 2$ and $\Lambda_{}^{\left(\text{opt}\right)}\left(d_{k1},d_{k0}\right)=\log P\left(\mathbf{r}_k|d_{k1},d_{k0}\right)$, we arrive at
\begin{equation}   \label{PLam}
P\left(\mathbf{r}_k|d_{k1},d_{k0}\right) \approx e^{\Lambda_{}\left(d_{k1},d_{k0}\right)-2N\log 2}.
\end{equation}
It is then clear that 
\begin{equation}
P\left(\mathbf{r}_k|d_{k0}=\pm 1\right)= \sum\limits_{d_{k1}\in \left\{+1,-1 \right\}} P\left(d_{k1}\right)P\left(\mathbf{r}_k|d_{k1},d_{k0}=\pm 1\right).
\end{equation}
Without considering the joint iteration between decoder and demodulator, we can assume that $P\left(d_{k1}=1\right)=P\left(d_{k1}=-1\right)$. Thus, the LLR of data $d_{k0}$ can be given as
\begin{equation}
\Lambda^{}\left(d_{k0}\right)=\log\frac{e^{\Lambda\left(d_{k1}=+1,d_{k0}=+1\right)} + e^{\Lambda\left(d_{k1}=-1,d_{k0}=+1\right)}} {e^{\Lambda\left(d_{k1}=+1,d_{k0}=-1\right)} + e^{\Lambda\left(d_{k1}=-1,d_{k0}=-1\right)}}.
\end{equation}
Noting that $\Lambda\left(1,1\right)=-\Lambda\left(-1,-1\right)$ and $\Lambda\left(1,-1\right)=-\Lambda\left(1,-1\right)$, we have
\begin{equation}\label{LLRF0}
\Lambda\left(d_{k0}\right)=\Lambda\left(d_{k1}=+1,d_{k0}=+1\right)+\Lambda\left(d_{k1}=-1,d_{k0}=+1\right).
\end{equation}
Similarly, the LLR of data $d_{k1}$ is given as
\begin{equation} \label{LLRF1}
\Lambda\left(d_{k1}\right)=\Lambda\left(d_{k1}=+1,d_{k0}=+1\right)+\Lambda\left(d_{k1}=+1,d_{k0}=-1\right).
\end{equation}
Substituting the estimated symbol log-likelihood functions given by (\ref{SubMonRv}) into (\ref{LLRF0}) and (\ref{LLRF1}), the LLRs of binary data $d_{k0}$ and $d_{k1}$ can be obtained from the monobit samples and training symbols.

\section{Monobit Receivers With IQ Imbalance}\label{RecImbalance}
In this section, we first investigate the effect of IQ imbalances. To mitigate the performance loss without increasing the complexity, monobit receivers based on double training sequences are then proposed. Finally, monobit receivers with eight-sector phase quantization is proposed to counter the IQ imbalance, at the cost of doubling the implementation complexity.

\subsection{Effect of IQ Imbalances}
We first examine the effect of amplitude imbalance. From (\ref{RecSigImb}) and (\ref{imbpara}), we arrive at 
\begin{equation}
\begin{split}
r_{\text{d},I}\left(t\right) &= \left(1+\alpha\right)\left[\cos\left(\theta/2\right)r_{\text{b},I}(t)+\sin (\theta/2)r_{\text{b},Q}(t)  \right]=(1+\alpha)r_{\text{d},I}^{(\alpha=0)}(t),\\
r_{\text{d},Q}(t) &= \left(1-\alpha\right)\left[\sin(\theta/2)r_{\text{b},I}(t)+\cos(\theta/2)r_{\text{b},Q}(t)  \right]=(1-\alpha)r_{\text{d},Q}^{(\alpha=0)}(t),
\end{split}
\end{equation}
where $r_{\text{d},I}^{(\alpha=0)}(t)$ and $r_{\text{d},Q}^{(\alpha=0)}(t)$ denote $r_{\text{d},I}(t)$ and $r_{\text{d},Q}(t)$ without amplitude imbalance, respectively. For practical systems, we also have $1+\alpha>0$ and $1-\alpha>0$. Let $r_{I,k,l}^{(\alpha=0)}$ and $r_{Q,k,l}^{(\alpha=0)}$ denote the corresponding samples of $r_{\text{d},I}^{(\alpha=0)}(t)$ and $r_{\text{d},Q}^{(\alpha=0)}(t)$ respectively. Thanks to the insensitiveness of monobit sampling to amplitude, we can immediately obtain
\begin{equation}\label{imbequ}
r_{I,k,l}=r_{I,k,l}^{(\alpha=0)}, ~\text{and}~~r_{Q,k,l}=r_{Q,k,l}^{(\alpha=0)}.
\end{equation}
This shows that the amplitude imbalance has essentially no impact on monobit receiver. However, a similar analysis shows that monobit receiver with eight-sector phase quantization is still affected by the amplitude imbalance.

We next discuss the impact of phase imbalance. When $\alpha=0$ and $\theta=0$, i.e. no IQ imbalance, we have
\begin{equation}  \label{IQEqu}
\begin{split}
r_{\text{p},I}\left(t|d_{k1}=1,d_{k0}=1\right) &=r_{\text{p},Q}\left(t|d_{k1}=1,d_{k0}=-1\right),\\
r_{\text{p},Q}\left(t|d_{k1}=1,d_{k0}=1\right) &=-r_{\text{p},I}\left(t|d_{k1}=1,d_{k0}=-1\right),
\end{split}
\end{equation}
where $r_{\text{p},I}\left(t|d_{k1},d_{k0}\right)$ and $r_{\text{p},Q}\left(t|d_{k1},d_{k0}\right)$ denote the signals of I and Q branches in noise-free channels respectively, when the symbol $\left(d_{k1},d_{k0}\right)$ is transmitted. However, (\ref{IQEqu}) does not hold when phase imbalance exists, rendering (\ref{OptMonR}) no longer optimal. Thus, the corresponding monobit receiver will suffer performance loss.

\begin{figure}
\centering
\includegraphics[width=2.5in]{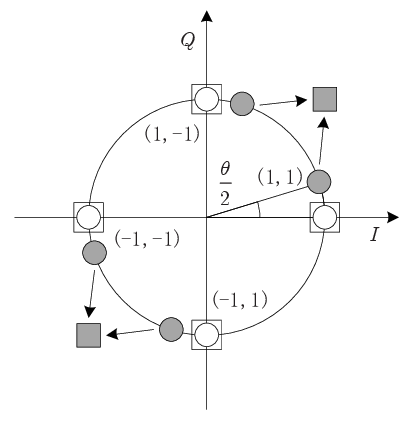}
\caption{Effect of phase imbalance in AWGN channel with high SNR when $\varphi=0^{\circ}$}
\label{fig.phaseimb}
\end{figure}

Take the AWGN channel for example. For the signals of one symbol duration, we have
\begin{equation}
\begin{split}
r_{\text{p},I}(t|1,1)&=(1+\alpha)p_\text{ref}(t)\cos(\varphi+\theta/2),\\
r_{\text{p},Q}(t|1,1)&=-(1-\alpha)p_\text{ref}(t)\sin(\varphi-\theta/2),\\
r_{\text{p},I}(t|1,-1)&=(1+\alpha)p_\text{ref}(t)\sin(\varphi+\theta/2),\\
r_{\text{p},Q}(t|1,-1)&=(1-\alpha)p_\text{ref}(t)\cos(\varphi-\theta/2),
\end{split}
\end{equation}
where $p_\text{ref}(t)$ is real. When $\rm sgn\left[\cos(\varphi+\theta/2)\right]=\rm sgn\left[ \cos(\varphi-\theta/2)\right]$ and $\rm sgn\left[\sin(\varphi+\theta/2)\right]=\rm sgn\left[ \sin(\varphi-\theta/2)\right]$, (\ref{IQEqu}) approximately holds and the impact of phase imbalance is weak due to the insensitiveness of monobit sampling, such as $\varphi=45^{\circ}$. Or else the reception performance will degrade significantly. One particularly severe situation is when phase offset $\varphi$ is around $0^{\circ}$ and SNR is high, as shown in Figure \ref{fig.phaseimb} where circles denote received symbols without quantization, rectangles denote the expectations of monobit quantization results of corresponding circles, and the gray and white denote the cases with and without IQ imbalances respectively. The amplitude imbalance $\alpha$ is set to be zero here for succinctness. It is observed that the monobit quantization expectations of different symbols are separate when phase imbalance is absent. However, the monobit quantization expectations of symbol $(1,1)$ and $(1,-1)$ tend to be the same when phase imbalance exists. The situation of $(-1,1)$ and $(-1,-1)$ is similar. In this case, the monobit receiver given by (\ref{SubMonRv}) will be confused and the reception performance is poor, especially when SNR is high. Hence, monobit receivers combating phase imbalance are desirable.

\subsection{Monobit Receivers Based on Double Training Sequences}
The noise of the I branch is no longer independent of the noise in the Q branch, when phase imbalance exists. Fortunately, such dependency is weak for practical values of $\theta$. Therefore, we assume they are independent in the sequel to simplify the analysis. Define
\begin{equation}
\begin{split}
&\epsilon_{I,l}^0=Q\left(r_{\text{p},I}\left(lT|1,1\right)\right),~~
\epsilon_{Q,l}^0=Q\left(r_{\text{p},Q}\left(lT|1,1\right)\right),\\
&\epsilon_{I,l}^1=Q\left(r_{\text{p},I}\left(lT|1,-1\right)\right),~~
\epsilon_{Q,l}^1=Q\left(r_{\text{p},Q}\left(lT|1,-1\right)\right).
\end{split}
\end{equation}
The log-likelihood function of the $k$th symbol with phase imbalance is given by
\begin{equation} 
\label{MLDbTr}
\begin{split}
\Lambda_\text{d}^{\left(\text{opt}\right)}\left(d_{k1},d_{k0}\right)=&\sum\limits_{i=I,Q}\sum\limits_{l=0}^{N-1} \left\{\log\left(1+\frac{d_{k1}+d_{k0}}{2} r_{i,k,l}\left(1-2\epsilon_{i,l}^0\right)\right)\right.\\
&\left. +  \log\left(1+\frac{d_{k1}-d_{k0}}{2} r_{i,k,l}\left(1-2\epsilon_{i,l}^1\right)\right)
\right\}-2N\log 2.
\end{split}
\end{equation}
Replacing $\Lambda_{}^{\left(\text{opt}\right)}\left(d_{k1},d_{k0}\right)$ in (\ref{MLDetector}) by $\Lambda_\text{d}^{\left(\text{opt}\right)}\left(d_{k1},d_{k0}\right)$, the ML monobit detector under phase imbalance is derived.

Define $w_{I,l}^0=1-2\epsilon_{I,l}^0$, $w_{Q,l}^0=1-2\epsilon_{Q,l}^0$, $w_{I,l}^1=1-2\epsilon_{I,l}^1$ and $w_{Q,l}^1=1-2\epsilon_{Q,l}^1$. Similar to the monobit ML receiver without phase imbalance, we perform two Taylor's expansions of (\ref{MLDbTr}), discard the constant $2N\log 2$, and obtain a suboptimal monobit receiver under phase imbalance as
\begin{equation}
\label{SubLLRDbTr}
\Lambda_\text{d}^{\left(\text{sub}\right)}\left(d_{k1},d_{k0}\right) = \sum\limits_{i=I,Q}\sum\limits_{l=0}^{N-1}\left\{\frac{d_{k1}+d_{k0}}{2} r_{i,k,l}\left(1-2\epsilon_{i,l}^0\right)+\frac{d_{k1}-d_{k0}}{2} r_{i,k,l}\left(1-2\epsilon_{i,l}^1\right)\right\}.
\end{equation}

To estimate the weights $w_{I,l}^0$, $w_{Q,l}^0$, $w_{I,l}^1$ and $w_{Q,l}^1$, we employ double training sequences. The first training sequence consists of $N_{t}^0$ symbols of $\left(1,1\right)$ and the second sequence consists of $N_{t}^1$ symbols of $\left(1,-1\right)$. Let $N_t=N_{t}^0+N_{t}^1$, so as to maintain the training cost. We may choose $N_{t}^0=N_{t}^1=N_t/2$ in practice. Thus, the weights can be estimated as
\begin{equation}
\label{EstiDbWgt}
\begin{split}
&\hat{w}_{I,l}^0 = \frac{1}{N_{t}^0}\sum\limits_{k=0}^{N_{t}^0-1}r_{I,k,l},~~~
\hat{w}_{Q,l}^0 = \frac{1}{N_{t}^0}\sum\limits_{k=0}^{N_{t}^0-1}r_{Q,k,l},\\
&\hat{w}_{I,l}^1 = \frac{1}{N_{t}^1}\sum\limits_{k=N_{t}^0}^{N_{t}-1}r_{I,k,l},~~~
\hat{w}_{Q,l}^1 = \frac{1}{N_{t}^1}\sum\limits_{k=N_{t}^0}^{N_{t}-1}r_{Q,k,l}.
\end{split}
\end{equation}
Substituting the estimated weights into (\ref{SubLLRDbTr}), the suboptimal monobit receiver under phase imbalance without CSI is obtained. 

To evaluate the performance of the monobit receiver given by (\ref{SubLLRDbTr}), its deflection ratio is calculated as
\begin{equation} 
\label{DefRatioDt}
D_\text{d}^{\left(\text{sub}\right)}=\frac{4\left[\sum\limits_{l=0}^{N-1}\left(1-2\epsilon_{I,l}^0\right) \left(\epsilon_{I,l}^1 - \epsilon_{I,l}^0\right) + \left(1-2\epsilon_{Q,l}^0\right)\left(\epsilon_{Q,l}^1- \epsilon_{Q,l}^0\right)\right]^2}
{\mathrm{var}\left(\lambda_\text{d}^{\left(\text{sub}\right)}\right)},
\end{equation}
where $\lambda_\text{d}^{\left(\text{sub}\right)}=\Lambda_\text{d}^{\left(\text{sub}\right)}\left(1,1\right)$ is the decision statistic and its variance is given by
\begin{equation}
\begin{split}
\mathrm{var}\left(\lambda_\text{d}^{\left(\text{sub}\right)}\right) &= \sum\limits_{l=0}^{N-1}\left\{  4\left[\epsilon_{I,l}^0\left(1-\epsilon_{I,l}^0\right)+\epsilon_{Q,l}^0\left(1-\epsilon_{Q,l}^0\right)\right]/N_t^0
+\left(w_{I,l}^0\right)^2 + \left(w_{Q,l}^0\right)^2 \right.\\
& \left.-0.5\left[ \left(w_{I,l}^0\right)^4+\left(w_{I,l}^0w_{I,l}^1\right)^2 + \left(w_{Q,l}^0\right)^4+ \left(w_{Q,l}^0w_{Q,l}^1\right)^2\right] \right\}.
\end{split}
\end{equation}
Analogously, the decision statistic can also be defined as $\lambda_\text{d}^{\left(\text{sub}\right)}=\Lambda_\text{d}^{\left(\text{sub}\right)}\left(1,-1\right)$ and the corresponding development is similar. It can be observed that if we increase both $N_t^0$ and $N_t^1$, the deflection ratio increases and the performance will improve.

To further boost the reception performance, we increase the equivalent number of training sequences without increasing training overhead. Since $\alpha$ and $\theta$ are usually small in practice, we may assume that
\begin{equation}  \label{iqcwapprox}
\begin{split}
&|r_{\text{p},I}\left(t|d_{k1}=1,d_{k0}=1\right)|\approx |r_{\text{p},Q}\left(t|d_{k1}=1,d_{k0}=-1\right)|,\\
&|r_{\text{p},Q}\left(t|d_{k1}=1,d_{k0}=1\right)|\approx |r_{\text{p},I}\left(t|d_{k1}=1,d_{k0}=-1\right)|.
\end{split}
\end{equation}
Considering monobit sampling is insensitive to amplitude, we may have
\begin{equation}
w_{I,l}^0\approx Aw_{Q,l}^1,~~w_{Q,l}^0\approx Bw_{I,l}^1,~~ A,B\in\{+1,-1\},
\end{equation}
where $A$ and $B$, called sign factors, represent the sign relationships between the weights. Due to the absence of the knowledge about $\alpha$, $\theta$, $\varphi$ and CSI, the sign factors need to be estimated from training sequences. Calculating the ML estimates of sign factors is extremely complex, so we propose a simple but effective estimation method as
\begin{equation}\label{signesti}
\hat{A} = {\rm sgn} \left(\sum\limits_{l=0}^{N-1}\hat{w}_{I,l}^0\hat{w}_{Q,l}^1\right),~~~
\hat{B} = {\rm sgn} \left(\sum\limits_{l=0}^{N-1}\hat{w}_{Q,l}^0\hat{w}_{I,l}^1\right).
\end{equation}
The numerical experiments show that such estimates are correct in most cases. Once the sign factors have been estimated, the weights estimated from training sequences can be combined as
\begin{equation} 
\label{WeightCW}
\hat{w}_{I,l}^{\left(\text{cw}\right)} = \frac{1}{2}\hat{w}_{I,l}^0+\frac{\hat{A}}{2}\hat{w}_{Q,l}^1,~~~
\hat{w}_{Q,l}^{\left(\text{cw}\right)} = \frac{1}{2}\hat{w}_{Q,l}^0+ \frac{\hat{B}}{2}\hat{w}_{I,l}^1~~.\end{equation}
It is observed that the variances of the combined weights are smaller than those of the original weights. Representing the original weights with the combined weights and substituting them into (\ref{SubLLRDbTr}), we obtain a monobit receiver as
\begin{equation}
\label{CWLLRDbTr}
\begin{split}
\Lambda_\text{d}^{\left(\text{cw}\right)}\left(d_{k1},d_{k0}\right) = \sum\limits_{l=0}^{N-1}\left\{\frac{d_{k1}+d_{k0}}{2}\left( \hat{w}_{I,l}^{\left(\text{cw}\right)}r_{I,k,l} + \hat{w}_{Q,l}^{\left(\text{cw}\right)}r_{Q,k,l}\right)\right.\\
\left. +\frac{d_{k1}-d_{k0}}{2}\left(\hat{B}\hat{w}_{Q,l}^{\left(\text{cw}\right)}r_{I,k,l}+ \hat{A}\hat{w}_{I,l}^{\left(\text{cw}\right)}\right)\right\}.
\end{split}
\end{equation}

To evaluate the possible performance gain offered by such receiver, its deflection ratio, denoted as $D_\text{d}^{\left(\text{cw}\right)}$, is calculated. The decision statistic is defined as $\lambda_k=\Lambda_\text{d}^{\left(\text{cw}\right)}\left(1,1\right)$. After manipulations, the numerator and denominator of $D_\text{d}^{\left(\text{cw}\right)}$ are given as
\begin{equation}
D_\text{d,num}^{\left(\text{cw}\right)} = \left[\sum_{l=0}^{N-1} \left(\epsilon_{I,l}^1-\epsilon_{I,l}^0\right)\left(w_{I,l}^0+Aw_{Q,l}^1\right)+ \left(\epsilon_{Q,l}^1-\epsilon_{Q,l}^0\right)\left(w_{Q,l}^0+Bw_{I,l}^1\right)\right]^2,  
\end{equation}
\begin{equation}
\begin{split}
D_\text{d,den}^{\left(\text{cw}\right)} &= \sum_{l=0}^{N-1} \left\{ \left[ V\left( \epsilon_{I,l}^0 \right) + V\left(\epsilon_{Q,l}^0\right) \right]/N_t^0 + \left[ V\left(\epsilon_{I,l}^1\right) +V\left(\epsilon_{Q,l}^1\right)\right]/N_t^1\right.\\
&\left.+0.5\left(w_{I,l}^0+Aw_{Q,l}^1 \right)^2 \left[ V\left(\epsilon_{I,l}^0\right) +V\left(\epsilon_{I,l}^1\right)\right]\right.\\
&\left.+0.5\left(w_{Q,l}^0+Bw_{I,l}^1\right)^2 \left[ V\left(\epsilon_{Q,l}^0\right) + V\left(\epsilon_{Q,l}^1\right)\right]\right\},
\end{split}
\end{equation}
and the deflection ratio is thus given by $D_\text{d}^{\left(\text{cw}\right)}= D_\text{d,num}^{\left(\text{cw}\right)}/D_\text{d,den}^{\left(\text{cw}\right)}$. Through numerical comparison, we find that $D_\text{d}^{\left(\text{cw}\right)}>D_\text{d}^{\left(\text{sub}\right)}$, which means that the receiver with combined weights can outperform the receiver given by (\ref{SubLLRDbTr}). 
This will also be shown in the numerical results in Section \ref{NumericalResults}.

\subsection{Monobit Receiver with Eight-Sector Phase Quantization}
To further improve monobit reception, more sophisticated strategies are needed. Compared with the monobit quantization under conventional receiver architecture, the eight-sector phase quantization scheme proposed in \cite{DBLP:journals/corr/abs-1112-4811} can provide more precise phase information of the received signal, at the price of two extra analog adders and monobit ADCs. By adding the I+Q and I-Q branches to the conventional monobit receiver, the eight-sector phase quantization can be implemented.

Define $\mathbf{\epsilon}_I=\left[\epsilon_{I,0},...,\epsilon_{I,N-1}\right]^T$ and $\mathbf{\epsilon}_Q=\left[\epsilon_{Q,0},...,\epsilon_{Q,N-1}\right]^T$. To simplify the notation, we let $\Lambda^{\left(\text{opt}\right)}\left(d_{k1},d_{k0}|\mathbf{\epsilon}_{I},\mathbf{\epsilon}_{Q}\right)= \Lambda^{\left(\text{opt}\right)}\left(d_{k1},d_{k0}\right)$ and $\Lambda^{}\left(d_{k1},d_{k0}|\mathbf{\epsilon}_{I},\mathbf{\epsilon}_{Q}\right)= \Lambda^{}\left(d_{k1},d_{k0}\right)$, where $\Lambda^{\left(\text{opt}\right)}\left(d_{k1},d_{k0}\right)$ and $\Lambda\left(d_{k1},d_{k0}\right)$ are given by (\ref{OptMonR}) and (\ref{SubMonRv}) respectively. If there is no IQ imbalance at the receiver, the optimal monobit receiver under eight-sector phase quantization is given by
\begin{equation}
\Lambda_\text{phq}^{\left(\text{opt}\right)}\left(d_{k1},d_{k0}\right) = \Lambda^{\left(\text{opt}\right)}\left(d_{k1},d_{k0}|\mathbf{\epsilon}_{I},\mathbf{\epsilon}_{Q}\right)+
\Lambda^{\left(\text{opt}\right)}\left(d_{k1},d_{k0}|\mathbf{\epsilon}_{-},\mathbf{\epsilon}_{+}\right), 
\end{equation}
where $\mathbf{\epsilon}_{-}$ and $\mathbf{\epsilon}_{+}$, defined similarly as $\mathbf{\epsilon}_I$ and $\mathbf{\epsilon}_Q$, are parameter vectors of the I-Q and I+Q branches respectively. The corresponding suboptimal monobit receiver can also be obtained as
\begin{equation}
\label{PhQSub}
\Lambda_\text{phq}^{\left(\text{sub}\right)}\left(d_{k1},d_{k0}\right) = \Lambda^{}\left(d_{k1},d_{k0}|\mathbf{\epsilon}_{I},\mathbf{\epsilon}_{Q}\right)+
\Lambda^{}\left(d_{k1},d_{k0}|\mathbf{\epsilon}_{I-},\mathbf{\epsilon}_{I+}\right) .
\end{equation}
For such receiver, the weights of all the four branches need to be estimated based on training sequences. Besides, iterative demodulation and removal of small-weight points are still useful.

When there are IQ imbalances at the receiver, the ML monobit receiver is much more complex. Thanks to the phase information offered by phase quantization, the effect of IQ imbalances is much weaker. As we will see in the numerical results in Section \ref{NumericalResults}, the suboptimal receiver proposed in (\ref{PhQSub}) is sufficient to combat the SNR loss caused by IQ imbalances.

\section{Numerical Results}\label{NumericalResults}
\subsection{Simulation Parameters}
The receivers obtained in this paper are particularly suitable to process wideband signals, such as those in UWB or 60GHz systems. In simulations, we use raised cosine pulse given as
\begin{equation}
p_\text{tr}\left(t\right)={\rm sinc}\left(t/\tau\right)\frac{\cos\left(\pi \beta t/\tau\right)}{1-4 \beta^2t^2/\tau^2},
\end{equation}
where $\tau$ is the time constant that controls the pulse duration, and $\beta$ is the roll-off factor.

In the following, simulation results are provided to evaluate the performance of the monobit receivers proposed in this paper. The raised cosine pulse was adopted with $\tau=0.5$ns and $\beta=1$. The carrier frequency was 4GHz. The bandwidth of the low-pass filter was $B$ = 5GHz, and the sampling rate was Nyquist rate $T$ = 100ps. We assume that there is no ISI and the timing is perfect. In AWGN channels, we set $T_s=10$ns and $N=100$. In multipath channels, we set $T_s=100$ns and $N=1000$ to avoid the ISI. The number of data symbols was $N_d=1000$. Considering the training overhead and demodulation latency, the length of training sequence was $N_t=100$ and the number of iterative demodulation was one. Thus, the training overhead amounted to 10 percent of the total transmission duration. The SNR is defined as $E_b/N_0=\sum_{l=0}^{N-1}p_\text{tr}^2\left(lT\right)$. For multipath channels, we used the aforementioned CM1 and CM14 channel models, both of which are used for indoor short-range communication where there is no high-speed moving object. Thus, the coherence time of these channels is dozens of milliseconds.

The overall BER performance was obtained by averaging those with different values of $\varphi\in\left[0^{\circ},90^{\circ}\right]$, which was sufficient due to the symmetry of phase offset. The step-size of $\varphi$ was set to be $3^{\circ}$. For each specific $\varphi$, the number of trials used in AWGN channel was 1000. For the multipath channels, both CM1 and CM14 channels were simulated with 100 realizations. The trials for each realization under each $\varphi$ was 100.

\subsection{Receiver Considered}
To simplify the notation in the next discussion, we use abbreviations. The first part of the abbreviation is either FR, MB or PQ, indicating whether full-resolution ADC, monobit sampling or eight-sector phase quantization is used. The second part is either E or F indicating whether estimated or perfect CSI is used in obtaining the weighting signal for detection. The third part is one of ML, MF, TE, indicating the type of weighting method used, corresponding to the optimal weights in (\ref{OptMonR}), the matched-filter weights, the suboptimal weights in (\ref{SubMonRv}) obtained from Taylor's expansion. For the simulations with IQ imbalances, the abbreviations DT and CW indicate the weights in (\ref{EstiDbWgt}) and (\ref{WeightCW}) respectively, based on double training sequences. Finally, the suffix IR indicates the iterative demodulation with removing small-weight samples, and SI indicates the sign factors are available at the receiver. Use such notation, the receivers that we will consider are as follows:
\subsubsection{FR-F-MF}
the optimal receiver with full-resolution sampling, perfect CSI, and matched-filter weights.
\subsubsection{MB-F-ML}
the optimal receiver with monobit sampling, perfect CSI, and optimal weights in (\ref{OptMonR}).       
\subsubsection{MB-F-MF}
the monobit receiver with perfect CSI and the matched filter weights.
\subsubsection{MB-F-TE}
the monobit receiver with perfect CSI, Taylor's expansion approximated weights.
\subsubsection{MB-E-TE}
the monobit receiver with estimated CSI, Taylor's expansion approximated weights.
\subsubsection{MB-E-TE-IR}
MB-E-TE receiver with removal of small-weight points and iterative demodulation.
\subsubsection{MB-F-MF-SI}
the monobit receiver with perfect CSI, the sign factor information, and matched filter weights.
\subsubsection{MB-E-DT}
the monobit receiver with estimated CSI, Taylor's expansion approximated weights based on double training sequences.
\subsubsection{MB-E-DT-IR}
MB-E-DT receiver with iterative demodulation and removal of small amplitude samples.
\subsubsection{MB-E-CW}
the monobit receiver with estimated CSI, combined weights based on double training sequences.
\subsubsection{MB-E-CW-IR}
MB-E-CW receiver with iterative demodulation and removal of small-weight points.
\subsubsection{PQ-F-TE}
the receiver with eight-sector phase quantization, perfect CSI, and suboptimal weights in (\ref{PhQSub}).
\subsubsection{PQ-E-TE-IR}
the receiver with eight-sector phase quantization, estimated CSI and suboptimal weights in (\ref{PhQSub}) with small-weight points removal and iterative demodulation.

\subsection{Numerical Results}

\begin{figure}
\centering
\includegraphics[width=\figwidth]{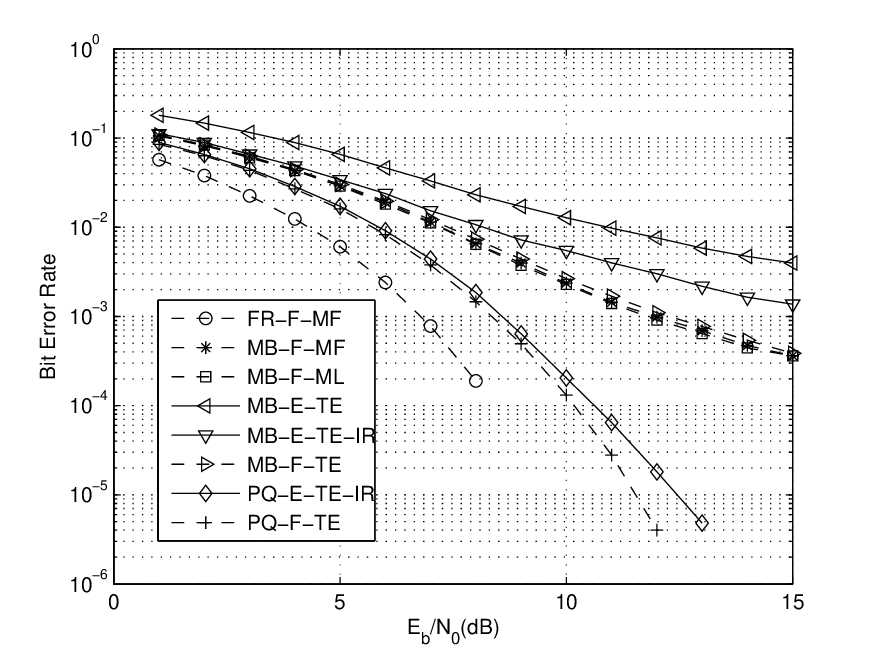}
\caption{Comparison of performance of optimal, suboptimal, and full-resolution, phase-quantization, monobit receivers in AWGN channel under Nyquist sampling without IQ imbalances}
\label{AWGNChannel}
\end{figure}

\begin{figure}
\centering
\includegraphics[width=\figwidth]{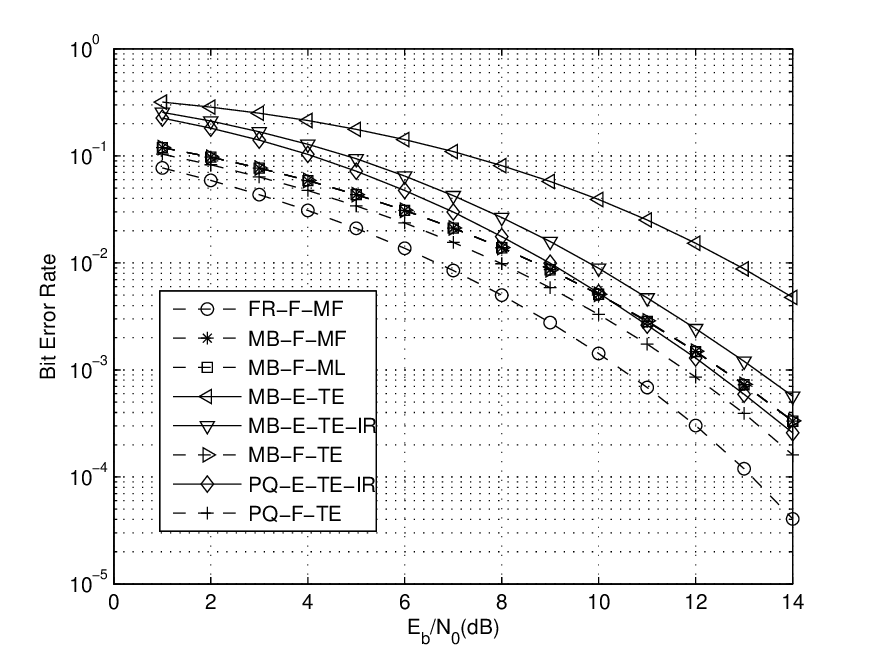}
\caption{Comparison of performance of optimal, suboptimal, and full-resolution, phase-quantization, monobit receivers in dense multipath channels under Nyquist sampling without IQ imbalances}
\label{CM1Perform}
\end{figure}

\begin{figure}
\centering
\includegraphics[width=\figwidth]{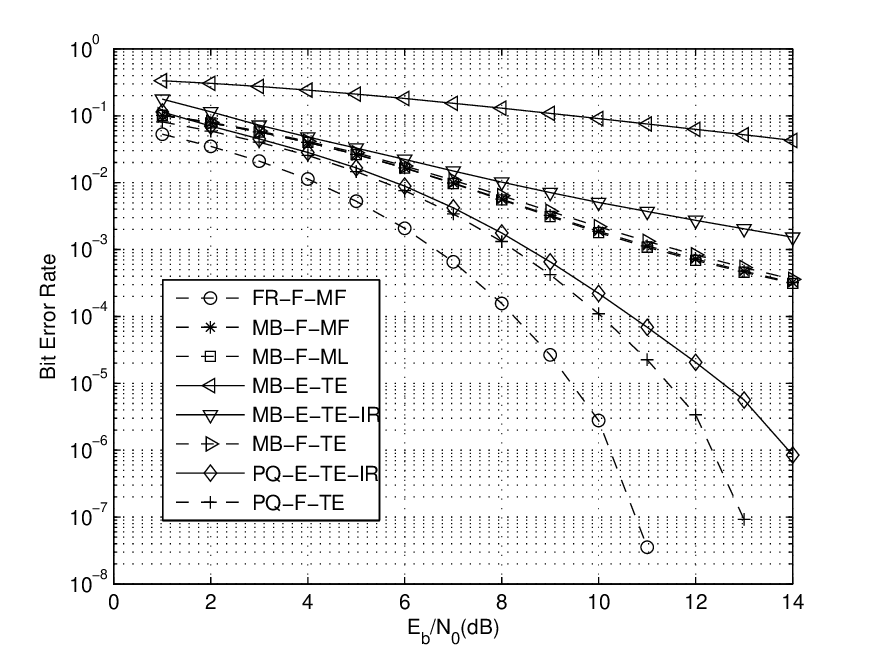}
\caption{Comparison of performance of optimal, suboptimal, and full-resolution, phase-quantization, monobit receivers in sparse multipath channels under Nyquist sampling without IQ imbalances}
\label{CM14Perform}
\end{figure}

Figure \ref{AWGNChannel} compares the performance of different receivers in AWGN channel without IQ imbalances. Given the perfect reference signal and phase offset (which is not possible in practice though), the MB-F-MF, MB-F-ML and MB-F-TE receivers have similar performance in entire SNR range. All of them have about 5dB SNR loss to the full-resolution matched filter when BER is around $10^{-3}$. Compared to the MB-E-TE receiver, the MB-E-TE-IR receiver can provide about 3dB performance gain, with about 2-3dB SNR gap from the monobit receiver with perfect CSI. It is shown that the performance loss of MB-E-TE-IR receiver in AWGN channel is mainly caused by the channel estimation error, which is unavoidable due to the highly nonlinear characteristic of the $Q$ function, especially in the low BER regime. By doubling the processing complexity in the digital domain (i.e. all operations in the digital domain, including weights estimation and demodulation computation, need to be done for four branches instead of two), the PQ-E-TE-IR receiver has 2dB SNR loss compared with the full-resolution matched filter. The impact of channel estimation error on PQ-E-TE-IR receiver is much weaker than MB-E-TE-IR receiver.

The performance under CM1 channel without ISI is shown in Figure \ref{CM1Perform}. Similar to AWGN channel, the MB-F-MF, MB-F-ML and MB-F-TE receivers have almost the same performance, which is about 2dB SNR loss to the FR-F-MF receiver. The MB-E-TE-IR receiver still outperforms the MB-E-TE receiver about 3dB performance gain, and has 3dB SNR loss even compared with the FR-F-MF receiver. It is also observed that the MB-E-TE-IR receiver can perform as well as the PQ-E-TE-IR receiver with less than 1dB SNR loss, thanks to the diversity of the dense multipath components and the phase diversity introduced by the multipath. We remark that such diversity gain is mainly contributed by a few paths with relatively large amplitudes, due to the removal of small amplitude samples, and thus it could still be available in the ISI case. The performance loss caused by channel estimation error is about 0.5dB, both for MB-E-TE-IR and PQ-E-TE-IR receivers. The performance of different receivers without ISI in CM14 channel is given in Figure \ref{CM14Perform}. In such sparse multipath channel, the PQ-E-TE-IR receiver can provide much better performance than the MB-E-TE-IR receiver, with double the complexity. It has less than 3dB SNR loss to the full-resolution matched filter when the BER is around $10^{-4}$.

\begin{figure}
\centering
\includegraphics[width=\figwidth]{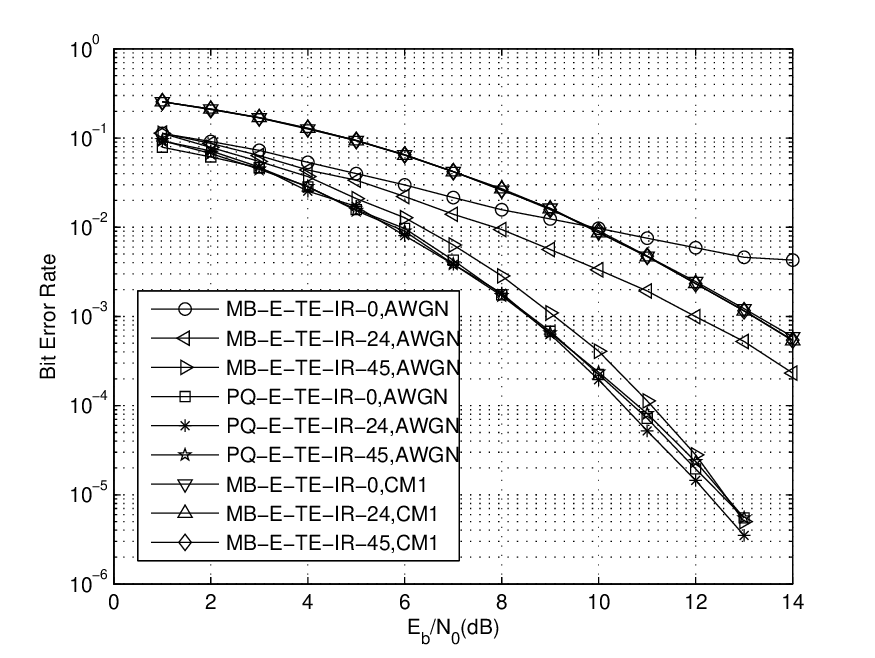}
\caption{Effect of phase offset on suboptimal monobit receivers under monobit, phase-quantization sampling without IQ imbalances}
\label{EffPhaseDif}
\end{figure}

\begin{figure}
\centering
\includegraphics[width=\figwidth]{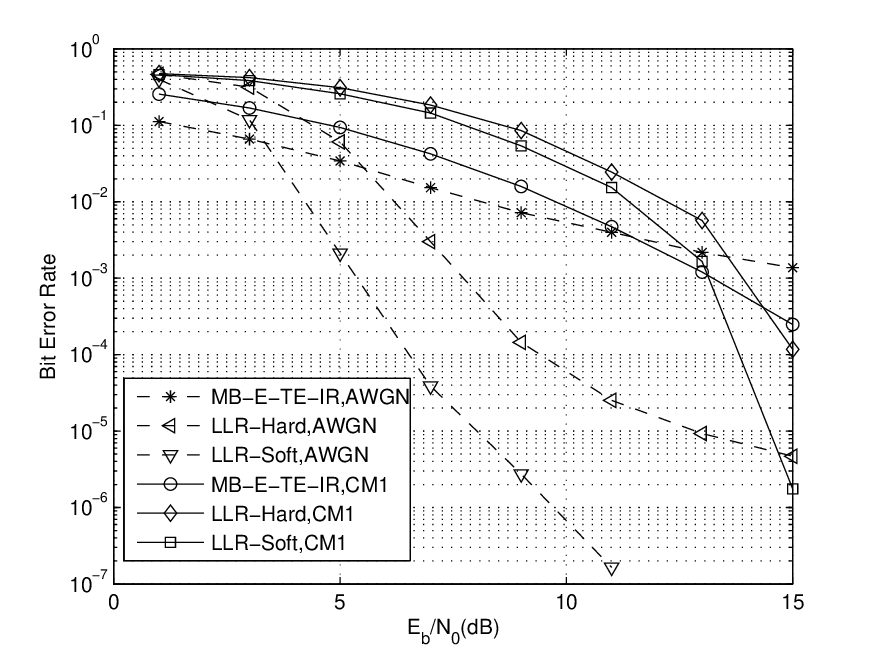}
\caption{Performance of MB-E-TE-IR receiver in coded setup without IQ imbalances}
\label{ErrInter}
\end{figure}

Figure \ref{EffPhaseDif} shows the effect of phase offset on the practical monobit receivers in AWGN and CM1 channels without IQ imbalances. The numbers suffixed to the abbreviations of the receivers indicate the degree of the phase offset, e.g. $\varphi=0^{\circ}$. It is observed that the MB-E-TE-IR receiver is greatly affected by the phase offset in AWGN channel. The reception performance is much better when the amplitudes of the I and Q branches are closer to each other ($\varphi=45^{\circ}$), which is consistent with the results derived from the deflection ratio in Figure \ref{DefRatioEps}. On the other hand, the PQ-E-TE-IR receiver can avoid being affected by the phase offset, due to the more precise phase information of the received signal it has. It also shows that the effect of phase offset on MB-E-TE-IR receiver in dense multipath channel is negligible.

Figure \ref{ErrInter} presents the performance of MB-E-TE-IR receiver without IQ imbalances in the coded case under different channel conditions. The typical 1/2 rate convolutional code with generator polynomial matrix of $\left[171,133\right]$ was employed. The abbreviations LLR-Hard and LLR-Soft indicate the hard and soft decoding respectively. It shows that the interface with soft decoding given in (\ref{LLRF0})-(\ref{LLRF1}) works well. It can offer much better performance than hard decoding, both in AWGN and CM1 channels. Considering its simplicity, it is a strong candidate for practical communication systems.

\begin{figure}
\centering
\includegraphics[width=\figwidth]{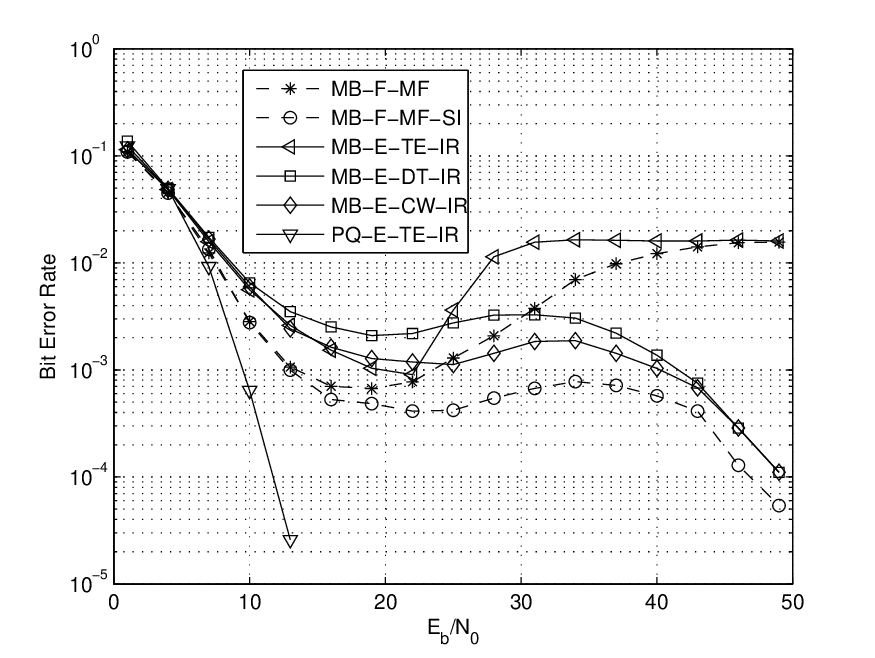}
\caption{Comparison of performance of different monobit receivers in AWGN channel with IQ imbalances}
\label{IMBAWGN}
\end{figure}

\begin{figure}
\centering
\includegraphics[width=\figwidth]{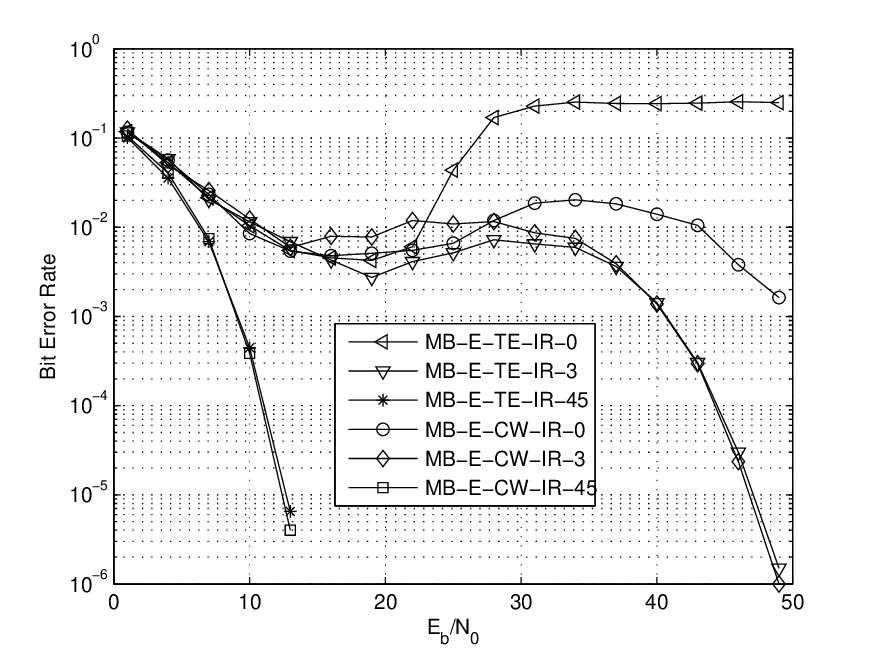}
\caption{Comparison of performance of different monobit receivers under different values of phase offset in AWGN channel with IQ imbalances}
\label{IMBAWGNPSI}
\end{figure}

\begin{figure}
\centering
\includegraphics[width=\figwidth]{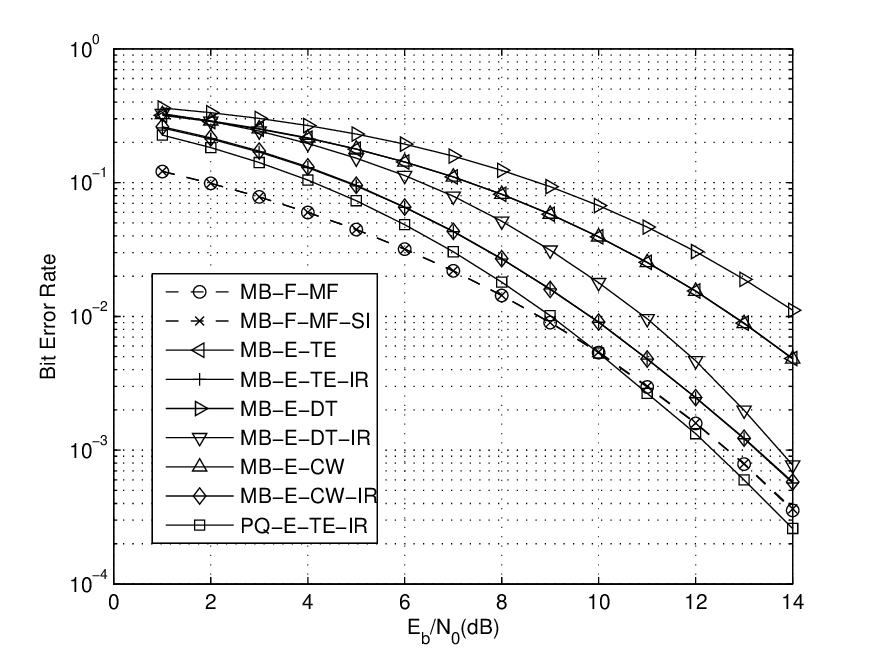}
\caption{Comparison of performance of different monobit receivers in dense multipath channels with IQ imbalances}
\label{IMBCM1}
\end{figure}

\begin{figure}
\centering
\includegraphics[width=\figwidth]{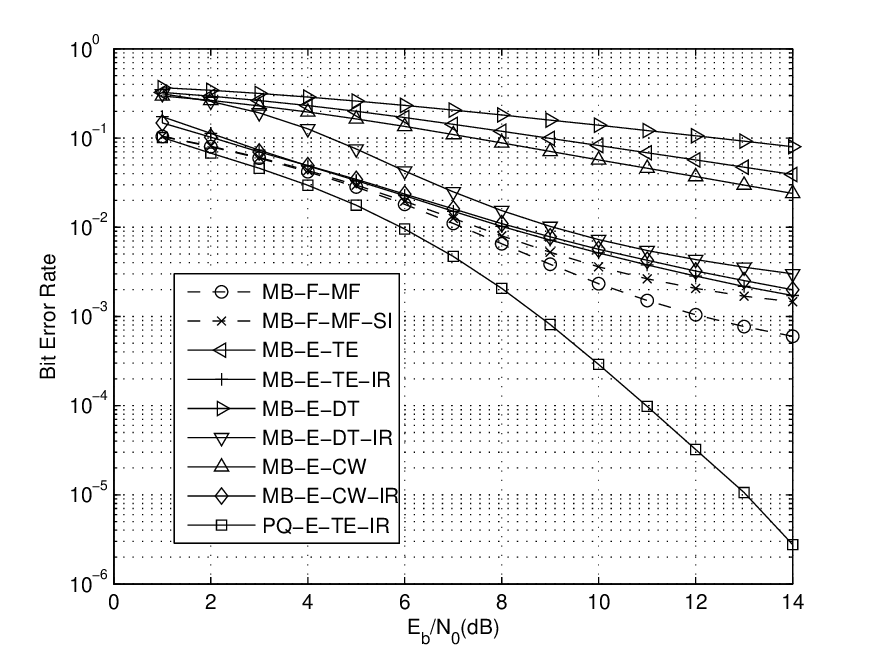}
\caption{Comparison of performance of different monobit receivers in sparse multipath channels with IQ imbalances}
\label{IMBCM14}
\end{figure}

Figure \ref{IMBAWGN} gives the performance of different receivers in AWGN channel with amplitude imbalance $\alpha=0.1$ and phase imbalance $\theta=5^{\circ}$. When double training sequences are used, we set $N_{t}^0=N_{t}^1=N_t/2=50$ to maintain training cost. It shows that monobit receiver without the information of sign factors, such as MB-F-MF or MB-E-TE-IR, has a $10^{-2}$ error floor at high SNR region. Such error floor is caused by the poor performance when phase offset is around $0^{\circ}$, as we show in Figure \ref{IMBAWGNPSI} where receivers under different values of phase offset with IQ imbalances is presented. It is observed that the BER of MB-E-TE-IR receiver with $\varphi=0^{\circ}$ is constantly about $2.5\times 10^{-1}$ when SNR is high. It is because the monobit quantization expectations of symbol $(1,1)$ and $(1,-1)$ tend to be the same, as shown in Figure \ref{fig.phaseimb}. Consequently, receivers without sign factors will confuse $(1,1)$ with $(1,-1)$, leading to high BER. This is the worst case. To avoid such performance loss, the sign factors are to be estimated. With the information of sign factors, the MB-E-DT-IR and MB-E-CW-IR receivers eliminate such error floor by providing one order of magnitude lower BER with $\varphi=0^{\circ}$. When phase offset $\varphi=45^{\circ}$, the performance of MB-E-CW-IR receiver is similar to that of MB-E-TE-IR receiver, thanks to the insensitiveness of monobit sampling to signal amplitude. It is also noted that both MB-E-DT-IR and MB-E-CW-IR receivers have a BER upturn when SNR is in 25-40dB. This is caused by the side-lobes of the pulse and the IQ imbalance. The MB-E-CW-IR receiver outperforms the MB-E-DT-IR receiver as analyzed before. It can also be observed that the PQ-E-TE-IR receiver can completely eliminate the effect of IQ imbalances.

Figure \ref{IMBCM1} presents the performance of different receivers with IQ imbalances in CM1 channels. The parameters of IQ imbalances are the same as those in AWGN channel. Thanks to the diversity offered by the dense multipath and the corresponding phase diversity, all receivers have almost no SNR loss compared with their corresponding performance without IQ imbalances in Figure \ref{CM1Perform}, except for the MB-E-DT-IR receiver whose performance is limited by the equivalent number of training sequences. The MB-E-CW-IR and MB-E-TE-IR receivers have almost the same performance. The PQ-E-TE-IR receiver has only about 1dB SNR gain to the MB-E-TE-IR or MB-E-CW-IR receiver, which perhaps is uneconomical compared to its complexity increasing. It is observed that the impact of IQ imbalances on monobit receivers in dense multipath channels is negligible. The performance of different receivers with IQ imbalances in CM14 channel is given in Figure \ref{IMBCM14}. In such sparse multipath channels, we observe that the PQ-E-TE-IR receiver can provide considerable performance gain. As a result, the trade-off between the performance and implementation complexity need to be made.

\section{Conclusions}\label{Conclusions}
We have obtained the optimal monobit receiver for QPSK and its performance in the form of deflection ratio. To reduce the implementation complexity, we proposed a suboptimal monobit receiver. We investigated the effect of phase offset between transmitter and receiver. The interface with error-control decoder was also derived. The simulation results showed that such receiver greatly reduces the complexity with about 3dB SNR loss in AWGN channel and only 1dB SNR loss in dense multipath channels, compared with matched-filter based monobit receiver with perfect CSI. 

We have also examined the impact of IQ imbalances at the receiver. Monobit receivers based on double training sequences are proposed to counter the performance loss caused by IQ imbalances, without increasing the complexity. Moreover, monobit receiver with eight-sector phase quantization is proposed to completely eliminate the effect of IQ imbalances. It is demonstrated that the amplitude imbalance has essentially no effect on monobit receivers. We noticed that the proposed monobit receiver can efficiently compensate for the SNR loss in AWGN channel, especially when SNR is high. The SNR loss of all these receivers in dense multipath channels is acceptable, thanks to the diversity offered by the multipath.

For cost and complexity consideration, the digital receivers with monobit sampling are strong candidates for future communication systems with significantly large bandwidths, such as UWB or 60GHz communications. There are several open issues to be addressed, such as evaluating the performance of monobit receiver under QAM modulation, the impact of IQ imbalances at the transmitter and the impact of timing imperfection.



\appendices


\section*{Acknowledgment}
The authors would like to thank the anonymous reviewers for their constructive comments on this work.

\ifCLASSOPTIONcaptionsoff
  \newpage
\fi

\end{document}